\documentclass[12pt]{article}
\usepackage{amsmath,amssymb,epsfig}
\usepackage{graphicx,floatflt,subfigure}
\usepackage{epstopdf}
\usepackage{color}
\input{colordvi.tex}
\usepackage[table]{xcolor}

\makeatletter

\renewcommand\section{\@startsection {section}{1}{\z@}%
                                   {-3.5ex \@plus -1ex \@minus -.2ex}%
                                   {2.3ex \@plus.2ex}%
                                   {\normalfont\large\bfseries}}

\renewcommand\subsection{\@startsection{subsection}{2}{\z@}%
                                     {-3.25ex\@plus -1ex \@minus -.2ex}%
                                     {1.5ex \@plus .2ex}%
                                     {\normalfont\normalsize\bfseries}}

\makeatother

\begin{document}

\baselineskip=18pt  
\numberwithin{equation}{section}  
\allowdisplaybreaks  



%
%


\thispagestyle{empty}

\vspace*{-2cm}
\begin{flushright}
\end{flushright}

\begin{flushright}
\end{flushright}

\begin{center}

\vspace{2.4cm}

{\bf\Large Catalytic Creation of Baby Bubble Universe }
\vspace*{0.2cm}

{\bf\Large with Small Positive Cosmological Constant}

\vspace{1.3cm}

{\bf
Issei Koga$^{1}$ and Yutaka Ookouchi$^{2,1}$} \\
\vspace*{0.5cm}

${ }^{1}${\it Department of Physics, Kyushu University, Fukuoka 819-0395, Japan  }\\
${ }^{2}${\it Faculty of Arts and Science, Kyushu University, Fukuoka 819-0395, Japan  }\\

\vspace*{0.5cm}

\vspace*{0.5cm}

\end{center}

\vspace{1cm} \centerline{\bf Abstract} \vspace*{0.5cm}

We investigate the decay of metastable de Sitter, Minkowski and anti-de Sitter vacua catalyzed by a black hole and a cloud of strings. We apply the method to the creation of the four dimensional bubble universe in the five dimensional anti-de Sitter spacetime recently proposed by Banerjee, Danielsson, Dibitetto, Giri and Schillo \cite{Danielsson}. We study the bounce action for the creation and find that the bubble with very small cosmological constant, of order $\Lambda^{(4)}/M^2_4 \sim 10^{-120}$, is favored by the catalysis by assuming appropriate mass scales of the black hole and the cloud of strings to reproduce the present energy densities of matter and radiation in the bubble universe.

\newpage
\setcounter{page}{1} 



\section{Introduction}

The discovery of Higgs particle and recent precise measurements of the top quark mass suggest that our universe may be metastable \cite{MetastableHiggs} as was firstly pointed out by \cite{Turner}. Metastability seems to be ubiquitous  in unified theories; In a supersymmetric grand unified theory, typically, there are several breaking patterns of a gauge group depending on a choice of the representation of a Higgs field. Incorporating supersymmetry breaking effects, one often find that degeneracy of vacua are resolved and some of the vacua become metastable. See \cite{SUSYbreaking}, e.g. for reviews. In string theory, various vacuum structures are involved and the diversity of vacua is allowed. This is known as the string landscape \cite{STland}. The string landscape was recently being  discussed from the viewpoint of the swampland program \cite{Swampland1,Swampland2}. One of the striking conjectures in the swampland program would be the no de Sitter conjecture \cite{dSconjecture1,GK,dSconjecture2,dSconjecture3} that does not allow a vacuum with a positive cosmological constant in string theories. It attracts wide attentions and related attempts have been done from various viewpoints \cite{SwamplandRefs,KKLTSwamp} (see \cite{review} for reviews). On the other hand, the swampland program has been boosting investigations of another realization of our universe in string theories. In particular, the authors of \cite{Danielsson} proposed a bubble universe with a positive cosmological constant in five dimensional anti-de Sitter (AdS) spacetime. Since the bubble is time-dependent, the model cleverly evades the conjectures and realizes a four dimensional de Sitter spacetime. Remarkably, the bubble is on the boundary between two AdS vacua, the four dimensional gravity can be localized on the bubble \cite{Danielsson} in the same spirit as the Randall-Sundrum scenario \cite{RS}. Moreover, by introducing a black hole and a cloud of strings \cite{Cloud}  in five dimensions, the authors of \cite{Danielsson} reproduced precise contributions of matter and radiation to the Friedmann equation in four dimensions.

In this paper, we would like to go a step further on this newly opened avenue and discuss the creation probability in terms of the catalysis. Here, we clam that the black hole and the cloud of strings in five dimensions, which were key ingredients to realize the realistic four dimensional bubble universe in \cite{Danielsson}, can act as seeds for the inhomogeneous nucleation of the bubble and claim that the creation rate is highly enhanced compared to the homogenous one \cite{Coleman,CDL}. The idea of the catalytic decay of metastable vacuum was firstly discussed in \cite{Pole} and later used in the context of phenomenological model building \cite{Pheno} and applied to decay processes of stringy metastable vacua \cite{Ookouchi}. The inhomogeneous vacuum decay in gravity theory was discussed in \cite{Hiscock,Gregory1401}\footnote{Vacuum decay by effects of a black hole or a compact-star was firstly discussed in \cite{Moss} and recently in \cite{Oshita}. }. We proceed computations of vacuum decay along the lines of the papers, especially, by exploiting the technique to deal with a singular bounce solution developed in \cite{Gregory1401}.

The organization of this paper is as follows. In the aim of studying a decay process in string theories, in section 2, we firstly apply the techniques developed in \cite{Gregory1401} on the singular bounce solutions to decay processes in various dimensions. In section 3, we study catalytic effects triggered by string clouds. Connecting two metrics of the string cloud \cite{Cloud}, we construct bounce solutions in various dimensions and estimate the decay rate. In section 4, armed with these studies, we investigate the catalytic creation of the bubble universe in AdS$_5$ and search for the most probable universe in this scenario. We address that the bubble with very small cosmological constant, of order $\Lambda^{(4)}/M^2_4 \sim 10^{-120}$, is favored by the catalysis, provided that appropriate scales of the black hole and the cloud of strings are present\footnote{$M_4=\sqrt{c\hbar/8\pi G_4}$ is the reduced Planck mass in four dimensions. Hereafter, we will set $\hbar$ and $c$ to $1$ except the section 4. }. The section 5 is devoted to conclusions and discussions. In appendix A, we generalize the study of Coleman and de Luccia (CDL) \cite{CDL} to various dimensions and describe the bounce actions in terms of the hypergeometric function.

\section{Black hole catalysis in $D$-dimensions}

In this section, we discuss the catalytic decay of a vacuum induced by a black hole. Recently, Gregory, Moss and Withers developed the method to compute the action for a singular bounce solution \cite{Gregory1401}. In the aim of application to string theory, we apply their method to the bounce action in $D$-dimensions and compute the decay rate, basically along the lines of \cite{Hiscock,Gregory1401,Gregory1503,Gregory}. In \cite{Gregory1503}, some results in $D$-dimensions were shown, so we briefly review them and show some of explicit new results.  

\subsection{General analysis in $D$-dimensions}

Since we below deal with a solution for a black hole and a cloud of strings with different energy scales and cosmological constants on each side of a bubble, we briefly outline basic formulae for $D$-dimensional Einstein gravity and the junction conditions for the metric \cite{Israel}. The Einstein equation in $D$-dimensions is 
\begin{eqnarray}
  \label{eq:1}
  R_{\mu\nu}-\frac{1}{2}Rg_{\mu\nu}+\Lambda^{(D)} g_{\mu\nu}=8\pi G_DT_{\mu\nu} ~,
\end{eqnarray}
where $G_D$ is the higher dimensional Newton constant and $\mu, \nu=0,\cdots D-1$. By taking the trace, the Ricci scalar is given by 
\begin{eqnarray}
  \label{eq:2}
 R=\frac{2}{2-D}\left(8\pi G_D T-D\Lambda^{(D)} \right)~.
\end{eqnarray}
We focus on the wall and consider $D-1$ dimensional subspace. The indices $i, j$ are for the subspace, $i,j=0,\cdots D-2$. The Einstein equation \eqref{eq:1} reduces to
\begin{eqnarray}
  \label{eq:3}
  R_{ij}  =8\pi G_D\left(T_{ij}+\frac{1}{2-D}\gamma_{ij}T\right)-\frac{2}{2-D}\gamma_{ij}\Lambda^{(D)}~,
\end{eqnarray}
where $\gamma_{ij}$ is the metric on the subspace. Here we define some quantities: The vector $n^{\mu}$ is the normal vector perpendicular to the surface and satisfy $1=g_{\mu \nu}n^{\mu}n^{\nu}$. ${e_{(i)}}^{\mu}$ is the differentiation of the coordinate in $D$-dimensions with respect to that of the surface. By these quantities, the metric can be represented as  $\gamma_{ij}=g_{\mu \nu} {e_{(i)}}^{\mu}{e_{(j)}}^{\nu}$. Introducing the extrinsic curvature
\begin{equation}
K_{ij}=g_{\mu \nu} {n^{\mu}}_{; \gamma}{e_{(i)}}^{\gamma} {e_{(j)}}^{\nu}~,
\end{equation}
we consider the integral of the curvature over the small interval. Since $R_{ij}$ and $T_{ij}$ have discontinuities across the wall, the integrals over the small interval yield finite results, 
\begin{eqnarray}
  \label{eq:5}
  \int^{\epsilon_w}_{-\epsilon_w}dl\, R_{ij}=K^+_{ij}-K^-_{ij}\ ,\qquad   \int^{\epsilon_w}_{-\epsilon_w}dl\, T_{ij}=S_{ij}~.
\end{eqnarray}
By putting all of these results together, \eqref{eq:3} becomes
\begin{eqnarray}
  \label{eq:4}
  K^+_{ij}-K^-_{ij}=8\pi G_D\left(S_{ij}-\frac{1}{D-2}\gamma_{ij}S\right)~.
\end{eqnarray}
Suppose that the energy momentum tensor is given by $S_{ij}=-\sigma \gamma_{ij}$. By taking the trace, we obtain $S=-(D-1)\sigma$ and the discontinuity of the extrinsic scalar curvature is given by 
\begin{eqnarray}
  K^+-K^-=8\pi G_D\frac{D-1}{D-2}\sigma~.\label{Israel}
\end{eqnarray}

Now we are ready to study an explicit metric in $D$-dimensions. We assume that the metric is of the following form, 
\begin{equation}
ds^2= -f(r) dt^2 +{dr^2 \over f(r)}+r^2d\Omega^2~, 
\end{equation}
where $d\Omega^2$ is the $D-2$ dimensional unit round metric and 
\begin{eqnarray}
f(r)&=&1-{2 \Lambda^{(D)} r^2\over (D-1)(D-2)}-{16\pi G_D M \over (D-2) A_{D-2} r^{D-3} }~.
\end{eqnarray}
Here, $A_{D-2}={2\pi^{D-1 \over 2}}/\Gamma({D-1\over 2})$ is the area of the $D-2$ dimensional unit sphere. Below, we consider a junction of two metrics with different mass scales and cosmological constants. So, we use the subscript $+$ ($-$) for quantities outside (inside) the wall. Adopting the same conventions as \cite{Gregory1503}, we define $\eta=\bar{\sigma}l$, $\bar{\sigma}={4\pi G_D \sigma /(D-2)}$ and 
\begin{equation}
l^2={(D-1)(D-2) \over 2\Delta \Lambda^{(D)}} \ ,\qquad \gamma= {4\bar{\sigma}l^2 \over 1+4\bar{\sigma}^2 l^2}\ ,\qquad \alpha^2 =1+{2\Lambda^{(D)}_- \gamma^2 \over (D-1)(D-2)}~, \label{gammaalpha}
\end{equation}
where $\Delta \Lambda^{(D)} =\Lambda^{(D)}_+-\Lambda^{(D)}_-$. 
Parametrizing the radius of the wall in terms of the proper time $\lambda$ for an observer on the wall as $r=R(\lambda)$, the metric induced on the wall takes the Friedmann-Lemaitre-Robertson-Walker form,
\begin{equation}
ds^2=-d\lambda^2 +R^2(\lambda) d\Omega^2~.
\end{equation}
Following Coleman and de Luccia \cite{Coleman,CDL}, we compute the Euclidian action for the bounce solution. To do that, we introduce the Euclidian time $\tau $ defined by the Wick rotation, $t=-i\tau$. Plugging the explicit form of the metric into \eqref{Israel}, we get the Israel's junction condition \cite{Israel}
\begin{eqnarray}
  \frac{1}{R}(f_+\dot \tau_+-f_-\dot \tau_-)=-\frac{8\pi G_D}{D-2}\sigma~,\label{JunctionCondition}
\end{eqnarray}
where the dot stands for the differentiation with respect to the proper time. From the normalization condition for the normal vector, we have 
\begin{eqnarray}
  \label{ap:2}
  f_\pm\dot \tau_\pm^2+\frac{\dot R^2}{f_\pm}=1~.
\end{eqnarray}
By solving in $\dot{R}^2$, the equation for describing the evolution of $R$ reduces to
\begin{equation}
\dot{R}^2=\bar{\sigma}^2 R^2 -\bar{f}+{(\Delta f)^2 \over 16 \bar{\sigma}^2 R^2}~,\label{Rdot2Ddim}
\end{equation}
where we defined $\bar f=(f_++f_-)/2$ and $\Delta f=f_+-f_-$. For the sake of simplicity, we normalize the variables as follows:
\begin{equation}
\tilde{R}={\alpha R\over \gamma}\  ,\quad \qquad \tilde{\lambda}={\alpha \lambda \over \gamma}\  ,\quad \qquad \tilde{\tau}={\alpha \tau \over \gamma} ~.\label{Normalize}
\end{equation}
With this notation, \eqref{Rdot2Ddim} becomes
\begin{equation}
\left({d \widetilde{R} \over d \tilde{\lambda} } \right)^2=1-\widetilde{R}^2-{k_1+2k_2\over \widetilde{R}^{D-3}}-{k_2^2 \over \widetilde{R}^{2(D-2)} }~, \label{tilRdot}
\end{equation}
where we defined
\begin{eqnarray}
k_1&=& {16\pi G_D \over (D-2)A_{D-2}} \left( {\alpha \over \gamma} \right)^{D-3} \left[ M_-+(1-\alpha) {\Delta M \over 2 \bar{\sigma} \gamma} \right] ~,\\
k_2&=&{16\pi G_D \over (D-2)A_{D-2}} \left( {\alpha \over \gamma} \right)^{D-2}{\Delta M \over 4\bar{\sigma}} ~.
\end{eqnarray}
and $\Delta M=M_+-M_-$.

Next, we move on to the computation of the on-shell action for the bounce solution. Since the period of the bounce solution does not agree with the periodicity determined by a horizon in general, hence we have to deal with a conical singularity on the solution. In the following, we discuss such contribution to the action by means of \cite{Gregory1401}. Suppose that there exist several singularities. In the vicinities of the singularities are referred to as ${\cal B}=\sum_i {\cal B}_i$. Decompose the manifold into two parts and write the action as $I= I_{ {\cal M}-{\cal B} } +I_{\cal B}$. Each part has a boundary, so we add the Gibbons-Hawking term to the action, 
\begin{eqnarray}
I_{ {\cal M}-{\cal B} }&=&-{1\over 16\pi G_D} \int_{{\cal M}-{\cal B}} R -\int_{{\cal M}-{\cal B}}{\cal L}_m+{1\over 8\pi G_D}  \int_{\partial ({\cal M}-{\cal B})}K ~, \\
I_{ {\cal B} }&=& -{1\over 16\pi G_D} \int_{\cal B} R -\int_{{\cal B}}{\cal L}_m+{1\over 8\pi G_D} \int_{\partial {\cal B}} K~\label{ISingu}. 
\end{eqnarray}
We further devide the action $ I_{ {\cal M}-{\cal B} }$ into three parts,
\begin{eqnarray}
  \label{eq:7}
  I=I_{\mathcal B}+I_-+I_++I_{\mathcal W}~,
\end{eqnarray}
where ${\cal W}$ stands for the contribution from the wall. The integral of the wall can be written in terms of the tension $\sigma$ as
\begin{eqnarray}
  I_{\mathcal W}=-\int_{\mathcal W}\mathcal L_m=\int_{\mathcal W}\sigma~.
\end{eqnarray}
By using the following decomposition of the scalar curvature,
\begin{equation}
R={}^{(D-1)}R-K^2+K_{ij}^2-2\nabla_i (u^i\nabla_j u^j)+2\nabla_j(u^i\nabla_i u^j) \label{Rdecomp}~,
\end{equation}
the action for each side on the wall can be given by 
\begin{eqnarray}
  I_\pm=-\frac{1}{8\pi G_D}\int_{\mathcal W}K_\pm+\frac{1}{8\pi G_D}\int_{\mathcal W}n_{\pm j}u^i\nabla_i u^j~,
\end{eqnarray}
where  $u^{i}$ is the differentiation of the coordinate with respect to the proper time. By plugging back in \eqref{eq:7} and use \eqref{Israel} and \eqref{JunctionCondition}, the total action can be given by
\begin{eqnarray}
  \label{eq:10}
  I&=&I_{\mathcal B}+\int_{\mathcal W}\sigma-\frac{D-1}{D-2}\int_{\mathcal W}\sigma-\frac{1}{16\pi G_D}\int_{\mathcal W}(f^\prime_+\dot\tau_+-f^\prime_-\dot\tau_-)~, \\
  &=&I_{\mathcal B}-\frac{1}{D-2}\int_{\mathcal W}\sigma-\frac{1}{16\pi G_D}\int_{\mathcal W}(f^\prime_+\dot\tau_+-f^\prime_-\dot\tau_-)~.
\end{eqnarray}
According to general study on contributions to the action from the singularities \cite{Gregory1401}, the integral \eqref{ISingu} can be expressed as  
\begin{equation}
I_{\cal B}=-{1\over 4G_D}\sum_i {\cal A}_i~,
\end{equation}
where ${\cal A}_i$ is the areas of the horizons at the singularities. By putting all together, let us compute the on-shell action,
\begin{eqnarray}
  \label{eq:8}
  I=-\frac{1}{4G_D}(\mathcal A_h+\mathcal A_c)\underbrace{-\frac{1}{D-2}\int_{\mathcal W}\sigma-\frac{1}{16\pi G_D}\int_{\mathcal W}(f^\prime_+\dot \tau_+-f^\prime_-\dot \tau_-)}_{\textcircled{\scriptsize 1}}~,
\end{eqnarray}
where 
\begin{eqnarray}
  \label{eq:9}
  \textcircled{\scriptsize 1}=\frac{A_{D-2}}{16\pi G_D}\int d\lambda\left(2R^{D-3}-2(D-1)G_DM_+\right)\dot\tau_+-\left(2R^{D-3}-2(D-1)G_DM_-\right)\dot \tau_-~.
\end{eqnarray}
Eventually, the bounce action is given by subtracting the action for the initial state, $I_0$, from the on-shell action for the bounce solution $I_B$,
 \begin{equation}
 B=I_{B}-I_{0} \label{subtraction}~.
 \end{equation}
It is hard to find an analytic result on this integral, hence we numerically estimate it for some fixed parameters below.

\subsection{Catalytic decay of de Sitter vacuum to Minkowski vacuum}

As an illustration, we numerically calculate the bounce action for the decay of de Sitter vacua to Minkowski vacua in various dimensions. In general, in the de Sitter vacua, there are two horizons which collide with each other when the mass of the black hole is large enough. We denote the critical value of the mass
\begin{equation}
M_N^{(D)}={(D-2)A_{D-2 }l^{D-3}\over 16\pi G_D}c^{(D)}~,
\end{equation}
where $c^{(D)}$ is the dimensionally different numerical factor\footnote{The explicit values are $c^{(4)}=\frac{2}{3\sqrt{3}}$, $c^{(5)}=\frac{1}{4}$, $c^{(6)}=\frac{6}{25}\sqrt{\frac{3}{5}}$, $c^{(7)}=\frac{4}{27}$, $ c^{(8)}={50\over 343}\sqrt{\frac{5}{7}}$, $c^{(9)}=\frac{27}{256}$, $c^{(10)}={686\over 19683}\sqrt{7}$.}. We will show the actions as functions of the ratio $M_+/M_N^{(D)}$ for fixed $\eta$. 

\begin{figure}[htbp]
    \centering
    \includegraphics[width=9cm]{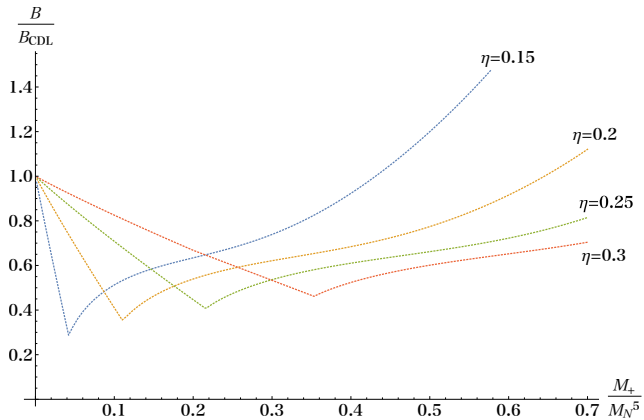}
    \caption{ Plots of the ratio $B/B_{CDL}$ as functions of the seed mass $M_+/M_N^{(5)}$ in five dimensions.}
    \label{FigA}
\end{figure}

We begin with the five dimensional case. Figure \ref{FigA} shows the bounce action normalized by that of Coleman-de Luccia discussed in the appendix.  For each choice of fixed $\eta$, the corresponding curve is constructed of two parts. One is the monotonically decreasing function reaching the minimal value at the critical point where the interval of the integral of the bounce action vanishes due to $R_{\rm min}=R_{\rm max}$. Hence, we refer to this bubble as the critical bubble and the corresponding seed mass as $M_{\rm crit}$. For an initial seed mass with $0\le M_+< M_{\rm crit}$, the dominant configuration corresponds to the nucleation of flat spacetime inside the bubble. For the case with $M_{\rm crit}\le M_+< M_{\rm max}$, the inhomogeneous decay is still dominant process for the decay but the ratio becomes an increasing function of the seed mass. The critical bubble with a black hole remnant gives a dominant contribution in this range. Finally, when the black hole mass is sufficiently large, we find that the inhomogeneous vacuum decay becomes sub-dominant compared to that of Caleman-de Luccia.

\begin{figure}[htbp]
    \centering
    \includegraphics[width=9cm]{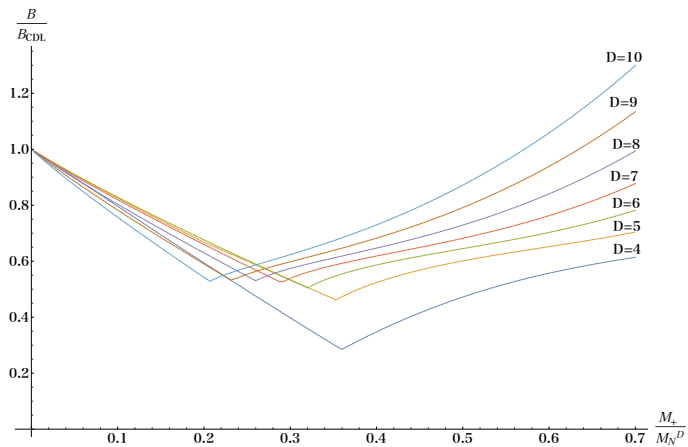}
    \caption{Plots of the minimum bounce action in various dimensions as functions of $M_+/M_N^{(D)}$ for fixed $\eta=0.3$.}
    \label{FigB}
\end{figure}

To see the dimensional dependence, we plot, in figure \ref{FigB}, the ratio $B/B_{ CDL}$ in various dimensions as functions of $M_+/M_N^{(D)}$. We find that the critical mass is a decreasing function in the dimension while the minimal value of the action for each critical point does not show any specific dependence on the dimensionality of spacetime. The lowest value of the bounce for $D=10$ is slightly lower than those for $D=8$ and $9$.

\section{String cloud catalysis}

In this section, we discuss the inhomogeneous vacuum decay in four and five dimensions caused purely by a cloud of strings. We focus on the decay of the de Sitter vacuum to the Minkowski vacuum. As for the decay of AdS vacua, we will treat in the next section. We claim that there is a different feature between four dimensional and five dimensional decays. In four dimensions, when the scale of the cloud of strings is sufficiently large, the semi-classical vacuum decay without the tunneling occurs.  

\subsection{General analysis}

For another seed of the catalysis, we consider a cloud of strings. It is constructed of relativistic strings \cite{Cloud}. In string theories and quantum field theories, there exist several origins for the cloud of strings such as a vortex solution generated by spontaneous symmetry breaking. Suppose that there is a spherically symmetric mass distribution which generates the Schwarzschild black hole geometry around it. If a number of strings are emanating from it, by smearing of the energy density, the geometry becomes the spherically symmetric cloud of strings discussed in \cite{Cloud}. We claim that the baryon vertex studied in \cite{Baryon} could be one such object in string theories. To be concrete, suppose that in the internal space, non-zero flux $F_5$ goes through a cycle $\mathbb{S}^5 $. From the Chern-Simons term in type IIB supergravity, we get the following low energy interaction,
\begin{equation}
\int_{10D}F_5\wedge B \wedge F_3= N \int_{5D} B\wedge F_3~.
\end{equation}
This is the BF coupling which is the universal low energy effective action for the discrete gauge theory. In this situation, when we wrap a D5-brane on $ \mathbb{S}^5 $, $N$-units of fundamental charges are induced on the brane, hence $N$ fundamental strings have to be attached to it. This object can play a role of the cloud of strings. 

The metric for the black hole and the cloud of strings in general dimension is given by \cite{Cloud}
\begin{equation}
ds^2= -f(r) dt^2 +{dr^2 \over f(r)}+r^2d\Omega^2~,
\end{equation}
where 
\begin{eqnarray}
f(r)&=&1-{2 \Lambda^{(D)} r^2\over (D-1)(D-2)}-{16\pi G_D M \over (D-2) A_{D-2} r^{D-3} } -{2a \over (D-2)r^{D-4}} \label{CloudMetric}~.
\end{eqnarray}
As in the previous section, consider a junction of two solutions with different scales. The equation of motion for the trajectory of the bubble becomes 
\begin{equation}
\left({d \widetilde{R} \over d \tilde{\lambda} } \right)^2=1-\widetilde{R}^2-{k_1+2k_2\over \widetilde{R}^{D-3}}-{k_2^2 \over \widetilde{R}^{2D-4} }  -{k_3+2k_4\over \widetilde{R}^{D-4}}-{k_4^2\over \widetilde{R}^{2D-6}}-{2k_2k_4\over \widetilde{R}^{2D-5}} \label{tilRdotCloud}~,
\end{equation}
where we defined
\begin{eqnarray}
k_3&=& {2\over  D-2}\left( {\alpha \over \gamma} \right)^{D-4} \left( a_-+{\Delta a (1-\alpha) \over 2\bar{\sigma}\gamma} \right)~, \\
k_4&=&{\Delta a \over  2\bar{\sigma}(D-2) }\left( {\alpha \over \gamma} \right)^{D-3 } ~,
\end{eqnarray}
and $\Delta a =a_+-a_-$. Solving the equation and plugging back into \eqref{subtraction}, we can compute the bounce action. Again, we numerically evaluate it since it is hard to find an analytic expression for the equation.

\subsection{String cloud catalysis in dS$_4$ spacetime}

Now, we are ready to study the first example of the catalysis induced by the cloud of strings. In this and next subsections, we demonstrate the catalysis without the black hole to elaborate effects of the cloud of strings. In section 4, we treat both seeds simultaneously. Suppose that an initial state having $\Lambda^{(4)}_+>0$ and $0<a_+<1$ decays to the Minkowski spacetime $\Lambda^{(4)}_-=0$. The equation of motion for $R$, in this case, is given by
\begin{equation}
\left( {d\widetilde{R} \over d \tilde{\lambda} }\right)^2=1-\left( \widetilde{R} +{k_4 \over \widetilde{R} } \right)^2-k_3~, 
\end{equation}
where 
\begin{equation}
k_3={a_-} \ , \qquad k_4={a_+-a_- \over 4\bar{\sigma}\gamma }~.
\end{equation}
For the sake of simplicity, we first discuss the bubble without a remnant, namely $a_-=0$.  The two solutions for the equation $\dot{\widetilde{R}}=0$ are given by
\begin{equation}
\widetilde{R}_{\rm min}={1\over 2} (1-\sqrt{1-4k_4})\ ,\qquad \widetilde{R}_{\rm max}={1\over 2} (1+\sqrt{1-4k_4}) ~.
\end{equation}
For the solutions to make sense, the condition $4k_4<1$ has to be satisfied. In other words, there is the upper limit for the cloud of strings for the tunnelling process, $a_+< {\eta \gamma / l}$. The initial state has the cosmological horizon,
\begin{equation}
f_+(r)=1-{\Lambda^{(4)}_+ r^2_c\over 3}-a_+=0\ , \qquad r_c=\sqrt{{3(1-a_+)\over \Lambda^{(4)}_+}}~.
\end{equation}
In computing the bounce action \eqref{subtraction}, the contribution coming from this horizon cancels out with the background action,
\begin{equation}
B=-{{\cal A}\over 4G_4}+I_a-( -{{\cal A}\over 4G_4})=I_a~,
\end{equation}
so the bounce action can be simply described by the following integral,
\begin{equation}
I_a= -{1\over 4G_4} \int d \lambda R^2\left[ \left(f_+^{\prime}-{2f_+\over R} \right)\dot{\tau}_+ - \left(f_-^{\prime}-{2f_-\over R} \right) \dot{\tau}_- \right]~.
\end{equation}

\begin{figure}[http]
\begin{center}
\includegraphics[width=.45\linewidth]{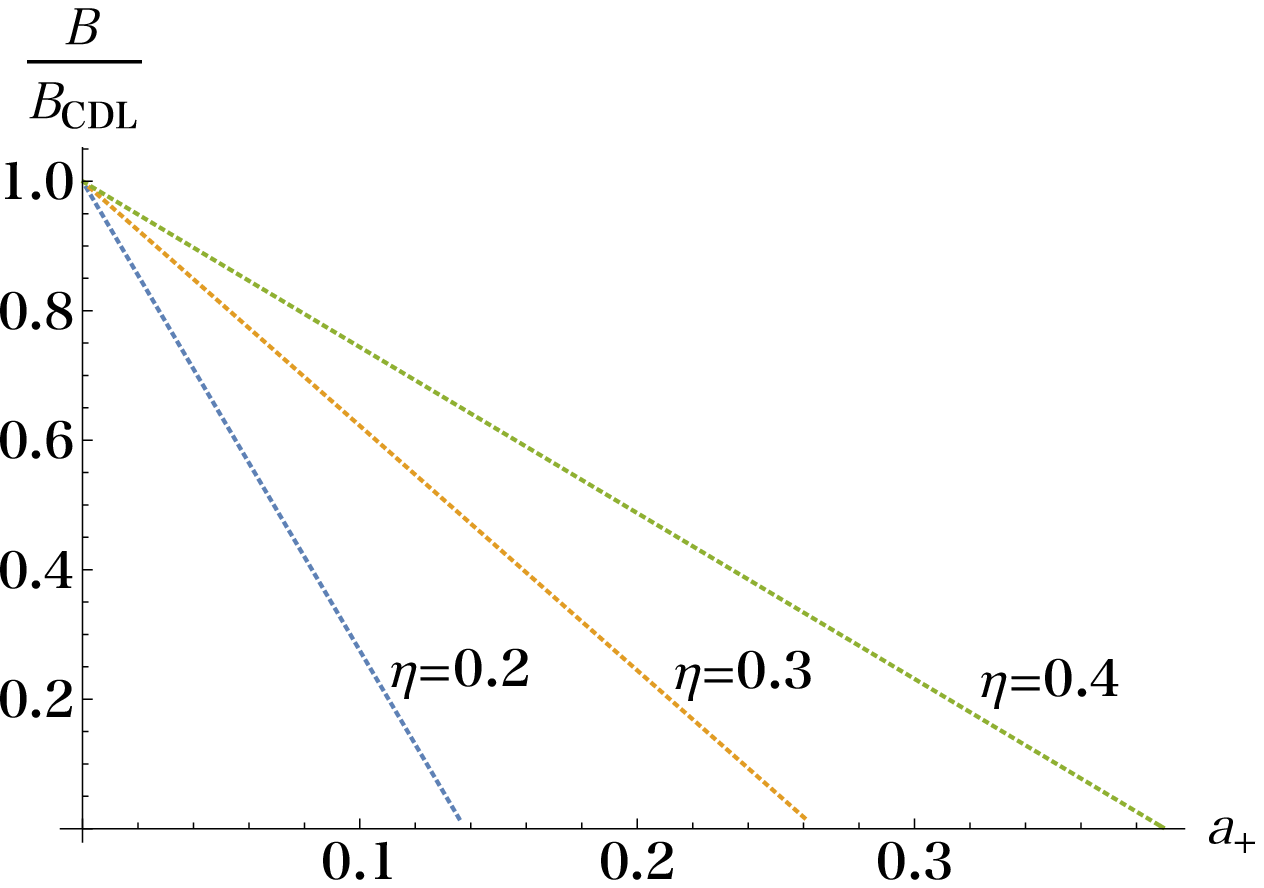}
\includegraphics[width=.45\linewidth]{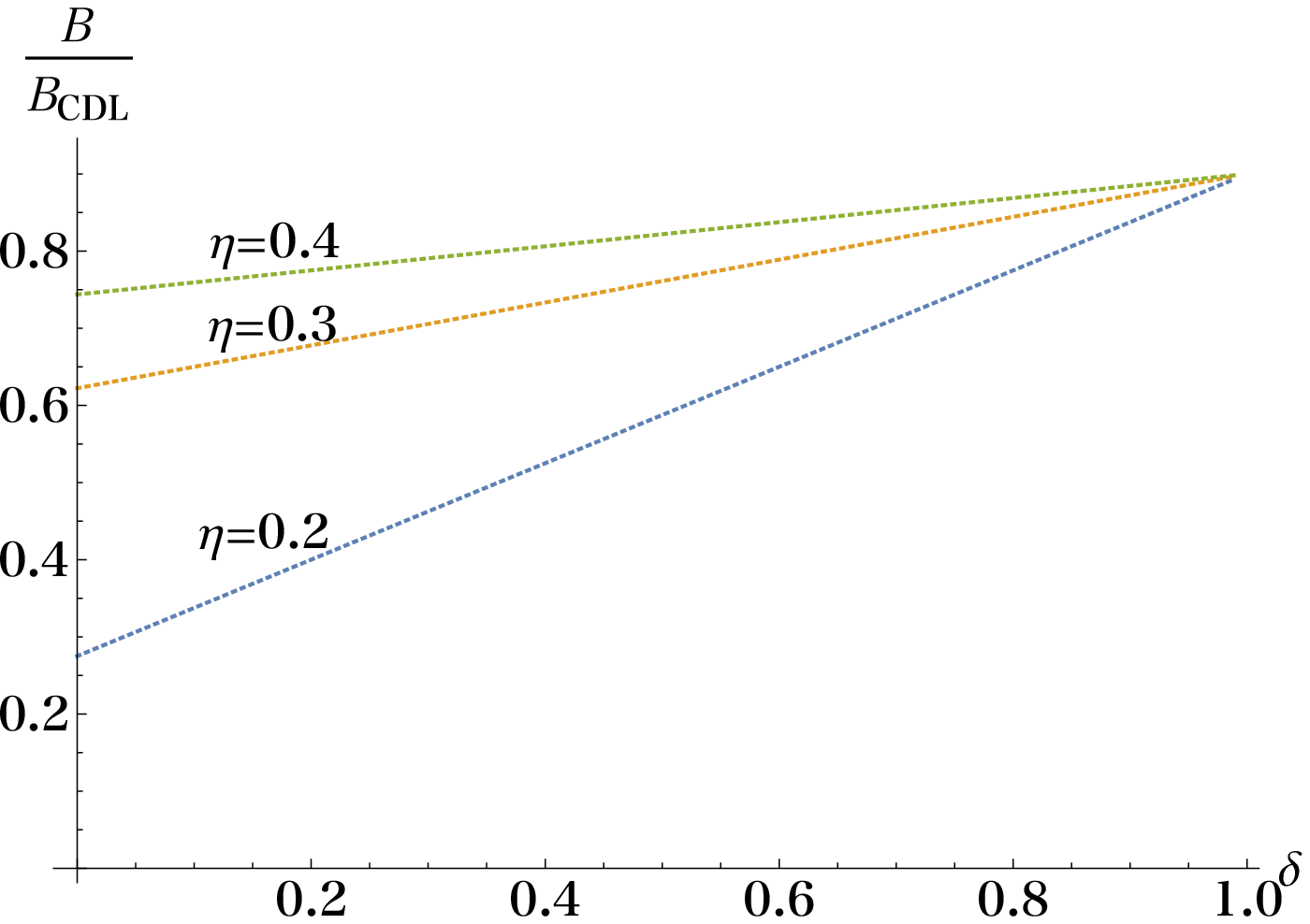}
\vspace{-.1cm}
\caption{\sl In the left panel, we show the bounce action for the bubble seeded by the cloud of strings without a remnant for the parameter choice  $\eta=0.2$, $0.3$ and $0.4$. In the right panel, the bounce action for fixed $a_+=0.1$ as a function of $\delta=a_-/a_+$ with $\eta=0.2$, $0.3$ and $0.4$. }
\label{FigC}
\end{center}
\end{figure}

As an illustration, we numerically calculate the bounce action in the figure \ref{FigC}. In the left panel, we show the action for the decay without a remnant for $\eta=0.2$, $0.3$ and $0.4$. The right panel shows the action with a remnant. From the first figure, we find that the bounce action monotonically decreases as $a_+$ becomes large until the critical value, $a^{\rm crit}_+=4\eta^2/(1+4\eta^2)$, above which the bounce action becomes zero. Thus, when the initial cloud is $a_+^{\rm crit}< a_+ <1$, the tunneling is not needed for vacuum decay, hence the semi-classical decay occurs, instead. This is remarkable because even if the vacuum itself is long-lived, the presence of the cloud of strings destabilizes it. This is in contrast to the results of the black hole catalysis. 

The difference comes from singularities of the bounce solution at horizons. In the case of black hole catalysis, there exist singularities which increase the energy-cost to construct the bounce configuration, which yields potential barrier to transit to the lower energy state. As the initial black hole mass gets larger,  more energy is required to create the bubble. This is the reason why the bounce action becomes an increasing function above the critical mass of the black hole in the figure 1 and 2. On the other hand, the string cloud solution does not have the event horizon, so when $R_{\rm min}$ approaches to $R_{\rm max}$, there is no extra energy-cost to generate the bounce configuration, which provides us the semi-classical vacuum decay.

\subsection{String cloud catalysis in dS$_5$ spacetime}

Here, we discuss the catalysis seeded by the cloud of strings in five dimensions. Again, we assume the de Sitter vacuum as the initial state and consider its decay to the Minkowski spacetime $\Lambda^{(5)}_-=0$. The equation of motion for $R$ is given by
\begin{equation}
\left( {d\widetilde{R} \over d \tilde{\lambda} }\right)^2=1-\widetilde{R}^2 -{k_3+2k_4 \over \widetilde{R}}   -{ k_4^2 \over \widetilde{R}^4}~.
\end{equation}
where we defined
\begin{equation}
k_3={2a_- \over 3\gamma}\equiv \widetilde{r}_- \ ,\qquad k_4={l\over 4 \eta \gamma}\left(  {2 a_+ \over 3 \gamma } -\widetilde{r}_-\right)~.
\end{equation}
In five dimensions, the string cloud solution has two horizons, as one can explicitly check from 
\begin{equation}
f_+(r)=1-{\Lambda^{(5)}_+ r^2\over 6}-{2a_+\over 3r}=0 ~. \label{F5Dcloud}
\end{equation}
These two horizons coincide with each other when the scale of the cloud of strings is $a_N={l/ \sqrt{3}}$. Also, in studying the bubble leaving a remnant after the transition, $a_-$ becomes nonzero. In this case, from the condition $f_-=1-{2a_- / 3r}=0$, we obtain the horizon that is given by $\widetilde{r}_-$ in dimensionless variable. 

\begin{figure}[htbp]
\begin{center}
\includegraphics[width=.55\linewidth]{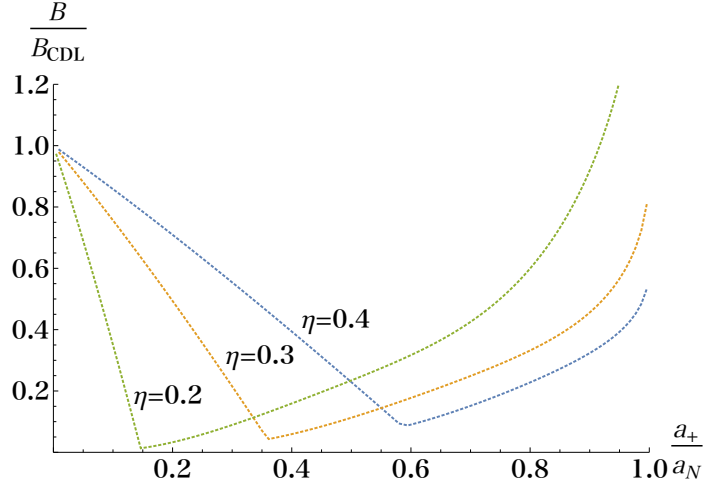}
\vspace{-.1cm}
\caption{\sl The bounce action for the catalytic decay of AdS${}_5$ to Minkowski vacuum.}
\label{FigE}
\end{center}
\end{figure}

In computing the bounce action, we should add contributions from the even horizons, hence, in total, the bounce action is given by  
\begin{equation}
B={r_+^3-r_-^3 \over 4G_5} +I_a~.
\end{equation}
We numerically evaluate this action for several choices of $\eta$ in the figure \ref{FigE}. From this, we see that there is the lower limit of the bounce action for each choice of $\eta$. This is due to contributions of the horizons, which is similar to the black hole catalysis.  Above this critical value of $a_+$, the bounce action becomes an increasing function as with the black hole catalysis. 

\section{Catalytic selection of cosmological constant}

The authors of \cite{Danielsson} proposed a new scenario to realize the four dimensional spacetime on a bubble separating two AdS$_5$ with different cosmological constants, which opens up a new avenue to construct a de Sitter spacetime in string theories. Here we would like to go a step further on this avenue and study the creation of a bubble in light of the catalytic effect induced by a black hole and a cloud of strings. In \cite{Danielsson,Kraus}, a black hole and a cloud of string are introduced to realize matter and radiation in the universe. We show that these ingredients can be seeds for inhomogeneous bubble nucleation and the tunneling probability is enhanced by the effect. 

We begin here with a review of the model. Consider a junction of two AdS$_5$ spacetimes. One has a negative cosmological constant $\Lambda_+^{(5)}$ and the other has a lower vacuum energy $\Lambda_-^{(5)}$, so that $|\Lambda_+^{(5)}|< |\Lambda_-^{(5)}|$. Plugging the explicit metric \eqref{CloudMetric} into the junction condition
\eqref{JunctionCondition}, the equation for describing the evolution of $R$ (in Minkowski-time) reduces to 
\begin{equation}
{\dot{R}^2\over R^2} \simeq -{1\over R^2}+{ \Lambda^{(4)}\over 3} +{8\pi \over 3}G_4 \left[{ M_+l_+-M_-l_- \over 2\pi^2 R^4}+{a_+ l_+-a_- l_-  \over  8\pi G_5R^3} \right]\label{Friedmann}~, 
\end{equation}
where we assume the late time evolution of the bubble with $R\gg l_{\pm}$ and $\dot{R}/R\gg l_{\pm}$. (Note that below, we will consider the early stage of the universe where its nucleation happens and the conditions,  $R\lesssim l_{\pm}$ and $\dot{R}/R\lesssim l_{\pm}$, are satisfied.)  Also, we defined
\begin{equation}
G_4={2 \over 1-\delta }{G_5\over l_+} \ ,\qquad l_{\pm}=\sqrt{-{6\over \Lambda^{(5)}_{\pm}}}\ , \qquad \delta ={l_-\over l_+} \label{G4G5}~.
\end{equation}
By introducing the parameter $\epsilon$ such that $\sigma= \sigma_{\rm crit}(1-\epsilon)$, where the critical value of the tension is defined by
\begin{equation}
\sigma_{\rm crit}={3\over 8\pi \delta l_+}{1 \over G_5}(1 -\delta)~,
\end{equation}
the cosmological constant in four dimensions can be identified with the quantities in five dimensions as follows,
\begin{equation}
  {\Lambda^{(4)} \over M_{4}^2}={96 \pi \hbar \over c^3}{G_5\over l_+^3} {\epsilon \over \delta ( 1-\delta ) }\label{CC} ~.
\end{equation}
For later reference, we show $\epsilon$ dependence on $\eta$, 
\begin{equation}
\eta \equiv {4\pi G_5 \over 3}\sigma l=\eta_{\rm crit}(1-\epsilon)\  ,\qquad \eta_{\rm crit}={1\over 2}\sqrt{1-\delta \over 1+\delta}~.
\end{equation}
The equation \eqref{Friedmann} is nothing but the Friedmann equation for the four dimensional spacetime with matter and radiation. If the tension of the bubble does not depend on time, the cosmological constant \eqref{CC} becomes constant literally. Since the bubble is expanding and time-dependent, this scenario does not contradict with the de Sitter conjecture \cite{dSconjecture1,GK,dSconjecture2,dSconjecture3} in string theories. From \eqref{Friedmann} we find that the black hole and the cloud of strings in five dimensions correspond to radiation and matter in four dimensions respectively. Remarkably, since the bubble exists on the boundary of two AdS$_5$ spacetimes, a zero-mode of the five dimensional graviton can be confined on the wall \cite{Danielsson} that gives rise to an effective four dimensional gravity in the same way as the Randall-Sundrum scenario \cite{RS}. 

\begin{figure}[htbp]
\begin{center}
\includegraphics[width=.6\linewidth]{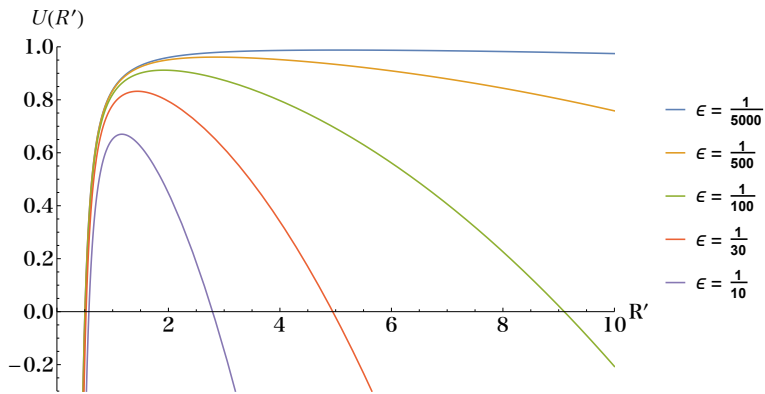}
\vspace{-.1cm}
\caption{\sl Plots of the effective potential $U(R^{\prime})$ with $8G_5M_+/3\pi l^2=4/100, \delta=2/10$ and $\widetilde{a}_+=1/10$. The curves from the bottom to the top correspond to $\epsilon=1/10$, $1/30$, $1/100$, $1/500$ and $1/5000$, respectively. }
\label{FigF}
\end{center}
\end{figure}

Now let us go back in time and study the very early stage of the universe where the catalytic creation of it is induced by the black hole and the cloud of strings. A difference from the previous section is non-existence of the cosmological horizon. In the present AdS$_5$ spacetimes there is no cosmological horizon, hence we can take large energy density of the seeds, which eventually enables us to realize very small positive cosmological constant in the four dimensions. By taking $\sigma$ is very close to the critical value $\sigma_{\rm crit}$, namely, when $\eta\simeq \eta_{\rm crit}$, $\epsilon$ becomes small, hence from \eqref{CC}, we can realize small cosmological constant. Ultimately, when $\eta$ is exactly equal to $\eta_{\rm crit}$, the parameter $\alpha$ vanishes, so the following different normalization of $R$ becomes useful,
\begin{equation}
\widetilde{R}= \alpha \widetilde{R}^{\prime}\ ,\quad  \qquad \widetilde{\lambda}= \alpha \widetilde{\lambda}^{\prime}\ ,\quad  \qquad \widetilde{\tau}= \alpha \widetilde{\tau}^{\prime}~.
\end{equation}
With these variables, the equation for the trajectory of the wall can be written as
\begin{equation}
\left( {d \widetilde{R}^{\prime} \over d \lambda^{\prime} }\right)^2=1-\alpha^2 \widetilde{R}^{\prime 2}-{k_1^{\prime}+2 \alpha k_2^{\prime} \over \widetilde{R}^{\prime D-3}}-{k_2^{\prime 2} \over \widetilde{R}^{\prime 2D-4}} -{k_3^{\prime}+2\alpha k_4^{\prime} \over \widetilde{R}^{\prime D-4}}-{k_4^{\prime 2} \over \widetilde{R}^{\prime 2D-6}}-{2k_2^{\prime }k_4^{\prime}\over \widetilde{R}^{\prime 2D-5}}\equiv U(R^{\prime})~,\nonumber
\end{equation}
where we defined
\begin{equation}
k_1=\alpha^{D-3}k_1^{\prime} \ , \qquad k_2=\alpha^{D-2}k_2^{\prime} \ , \qquad k_3=\alpha^{D-4}k_3^{\prime} \ , \qquad k_4=\alpha^{D-3}k_1^{\prime} ~. \label{finetune}
\end{equation}
We plot the potential $U(R^{\prime})$ in the figure \ref{FigF} for the parameters $\epsilon=1/10$, $1/30$, $1/100$, $1/500$ and $1/5000$. When $\eta$ approaches to the critical value, the width of the potential gets large enough, prohibiting the tunnelling process. In this case, in order to decay the vacuum efficiently, the catalytic effect becomes highly important.  

\begin{figure}[htbp]
\begin{center}
\includegraphics[width=.56\linewidth]{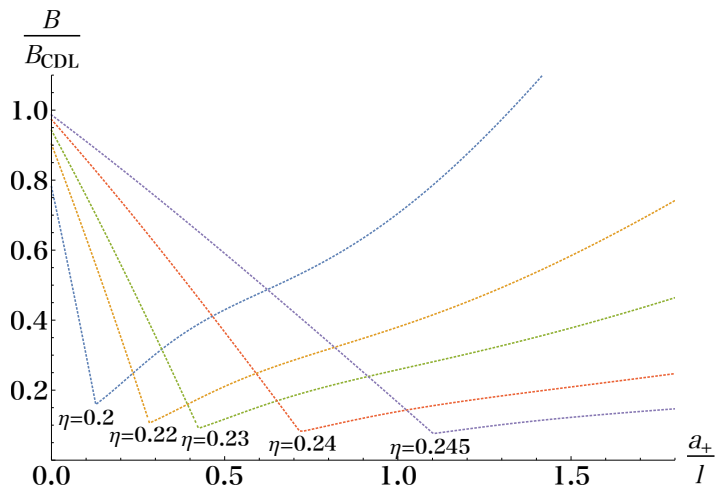}
\vspace{-.1cm}
\caption{\sl The bounce action for the decay of AdS${}_5$ to AdS${}_5$ with lower vacuum energy with $\eta=0.2$, $0.22$, $0.23$, $0.24$ and $0.245$. We chose $\delta={6/10}$, $8G_5M_+/3\pi l^2=4/100$.}
\label{FigG}
\end{center}
\end{figure}

Following the same procedure as before, we compute the bounce action in this scenario. We assume the decay of AdS$_5$ spacetime to another AdS$_5$ vacuum with a lower but the same order energy density. As an illustration, we show numerical results for the bounce action for the parameters $\delta=6/10$ and $8G_5M_+/3\pi l^2=4/100$ in the figure \ref{FigG}. For sufficiently small $a_+$, the bubbles without remnants $(M_-, a_-)=(0,0)$ are preferable and as $a_+$ increases, the bounce becomes small until the critical value above which a remnant remains. As in the black hole catalysis, the contributions from a singularity of the bounce solution yields the minimal value of the bounce action. 

For a realistic scenario, let us estimate the scales of the black hole and the cloud of strings by using the late time evolution of the bubble \eqref{Friedmann} as a boundary condition. By substituting the energy densities of matter and radiation at the present age for \eqref{Friedmann}, these scales can be determined as 
\begin{equation}
\widetilde{a}_+={a_+ \over l}={8\pi G_5 R^3 \rho^{\rm mat} \over c^2 l_+ l} \ , \qquad \beta={8G_5 M_+ \over 3\pi c^2 l^2}={16\pi G_5 R^4 \rho^{\rm rad} \over 3c^2 l^2l_+}~,
\end{equation}
where we assume $a_-=0$, $M_-=0$ because we are interested in the minimal value of the bounce action for fixed $\eta$. We take the Hubble horizon $R_0$ as the size of the bubble at the present age,
\begin{equation}
R_0=cH_0^{-1}\simeq 1.2\times 10^{26} [m] ~,
\end{equation}
and the energy densities of radiation and matter are given by 
\begin{equation}
\rho^{\rm rad}\simeq 7.3 \times 10^{-31} \left[ {kg \over m^3} \right] \ , \qquad \rho^{\rm mat}\simeq 2.3 \times 10^{-27} \left[ {kg \over m^3} \right]~.
\end{equation}
Using these values and $\delta=6/10$, one can estimate the scales of the black hole and the cloud of strings as follows:
\begin{equation}
\widetilde{a}^{(0)}_+\simeq {1.4\over \zeta_0}\times 10^{60}\ , \quad \qquad \beta^{(0)}\simeq{3.2\over \zeta_0^2} \times 10^{117} ~,\label{present}
\end{equation}
where we defined $l_+=\zeta_0 \sqrt{\hbar G_4/c^3 }$ by the tunable parameter $\zeta_0$ and used \eqref{G4G5}. 

\begin{figure}[htbp]
\begin{center}
\includegraphics[width=.42\linewidth]{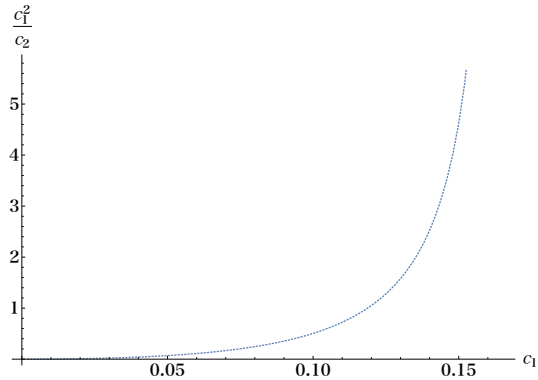}
\vspace{-.1cm}
\caption{\sl Numerical values of $c_1$ and $c_1^2/c_2$ for the choice $\delta=6/10$. }
\label{FigH}
\end{center}
\end{figure}

Next, we study small $\epsilon$ parameter range and examine the bounce action associated with the most efficient decay process which is the critical bubble without remnants. From the conditions, $\dot{\widetilde{R}}=d \dot{\widetilde{R}}/d \widetilde{R}=0$ and $(M_-,a_-)=(0,0)$, we find that 
\begin{equation}
\widetilde{a}^{(c)}_+\simeq {c_1\over \sqrt{\epsilon}} \ , \qquad \beta^{(c)}\simeq {c_2 \over \epsilon} ~, \label{CriticalValue}
\end{equation}
at the leading order in $\epsilon$. The values of $c_1$ and $c_2$ are shown in the figure \ref{FigH}. 

\begin{figure}[htbp]
\begin{center}
\includegraphics[width=.5\linewidth]{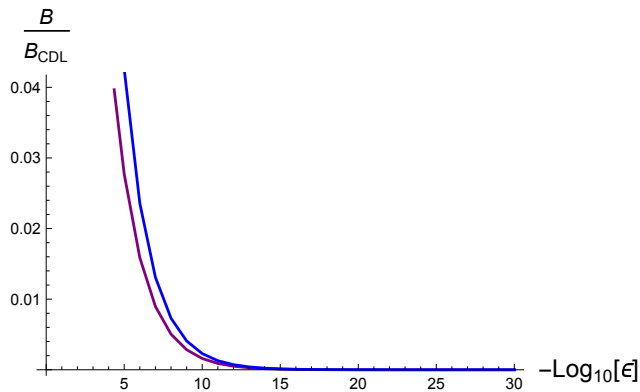}
\vspace{-.1cm}
\caption{\sl The ratio $B/B_{CDL}$ as a function of $\epsilon$ for fixed $\delta=6/10$. The purple and blue curves correspond to $c_1=1/100$ and $c_1=1/10$. }
\label{FigI}
\end{center}
\end{figure}

The values of \eqref{CriticalValue} yield the most dominant catalytic effect for the decay.  Comparing with the present value \eqref{present}, we determine the value of $\epsilon$. When the catalysis seeded by the cloud of strings dominates over that of the black hole, from \eqref{present} and \eqref{CriticalValue}, we obtain $\epsilon= 5.1\times \zeta^2_0 c_1^2 \times 10^{-121} $. With this value, we find that the ratio $\beta^{(0)}$ and $\beta^{(c)}$ , 
\begin{equation}
{\beta^{(0)} \over \beta^{(c)}}\simeq {1.6 c_1^2 \over c_2}\times 10^{-3}~,
\end{equation}
is very small when $c_1<0.15$, which indicates that the catalysis by the black hole does not work well\footnote{To the contrary, if we use the choice of $\epsilon$ determined by $\beta^{(0)}=\beta^{(c)}$  and estimate $\widetilde{a}_+^{(c)}$, we find that $\widetilde{a}_+^{(0)}/\widetilde{a}_+^{(c)}\gg 1$. This means that the scale of the cloud of strings is too large and the bounce action becomes larger than that of Coleman-de Luccia. Thus, the catalytic decay does not occur.}. With this value of $\epsilon$, from \eqref{CC} we find that  
\begin{equation}
{\Lambda^{(4)}\over M_4^2}={48\pi  \epsilon \over \delta l_+^2} {\hbar G_4 \over c^3}\simeq {1.2\times c_1^2}\times 10^{-118}=1.2\times 10^{-120}~,
\end{equation}
where at the last step we took $c_1=1/10$. This number is remarkable and reproduces the cosmological constant of our universe. We simply assumed the energy densities of radiation and matter at the present age as the input conditions, and studied the catalytic effect induced by the cloud of strings, then eventually we arrived at the precise order of the cosmological constant. Moreover, from the figure \ref{FigG}, we see that when $\eta$ approaches to $\eta_{\rm crit}$, the bounce action for the critical bubble without remnants becomes smaller. That indicates that among the various choices, small $\epsilon$ is preferable from the point of view of the catalytic decay. In fact, when $\epsilon$ is very small, the bounce action for the critical bubble without remnants is estimated as in the figure \ref{FigI}. Since it almost vanishes, the decay process is semi-classical, namely quantum tunneling is not required. 

Although we ``explained'' the order of the present cosmological constant by the catalysis in the context of the bubble universe proposed in \cite{Danielsson}, we should be careful on the cosmic history of the universe. In our discussion, this scenario does not incorporate the inflation at the early stage of the universe which can ruin the success of our argument. Also, the thermal history of the universe contributes to the calculations. Thus, it would be important to construct a realistic model and check if the idea of ``the catalytic selection of the cosmological constant'' works in the model.

Finally, we comment on tuning of the bubble tension $\sigma$ (equivalently $\eta$). Roughly, it depends on the difference between two cosmological constants and the shape of the potential in between such as the height and the width. Furthermore, it also depends on a path of the tunnelling. In the homogeneous bubble nucleation, the most efficient path determines the tension. Away from the minimum path, the tension becomes larger. Since there are infinite numbers of such non-minimal paths, the allowed bubble tension can have a large parameter space. As emphasized, one of the striking features of the catalytic decay of AdS vacua can be seen in the minimal value of the bounce action for each $\eta$ in figure \ref{FigG}. When $a_+$ is large, the dominant bubble does not correspond to the minimal value of $\eta$. For example, when $a_+/l \sim 0.5$, the bounce actions for $\eta=0.2$ and $0.22$ are larger than that of $\eta=0.23$. As mentioned above, it would be easy to increase the tension by taking a non-minimum path for the tunneling. Our argument suggests that during the phase transition, the most economical tension can be automatically selected by the catalysis. Moreover, if the string landscape is true, there can exist a large number of metastable anti-de Sitter vacua, which yields large tunable parameter space for the decay, hence the catalytic selection of cosmological constant would work well. We believe that this new selection mechanism of the cosmological constant adds a virtue to the bubble universe proposed in \cite{Danielsson} and provide a support for the scenario.

\section{Conclusions and discussions}

In this paper, we considered the decay of metastable vacua including gravitational effects. In particular, we focused on the inhomogeneous vacuum decay triggered by black holes and string clouds. Calculating the bounce action by exploiting the techniques developed in \cite{Gregory1401,Gregory1503,Gregory}, we searched for a parameter space where the catalytic decay dominants. From these results, we read off the tendency of the most preferable bubble. In general, for sufficiently low energy densities of the seeds, vacuum decay without remnants is preferable. In four dimensional catalysis by the cloud of strings, we find essentially different behavior from that of the black hole. That is the semi-classical vacuum decay; In increasing the initial energy density of the cloud of strings, we found the critical value above which the bounce action becomes zero. Hence, the tunneling is not required for vacuum decay. As for the BH catalysis, there were contributions coming from singularities of the bounce solution, giving a lower bound on the action. However, the cloud solution in four dimensions does not have the event horizon, so there is no minimal bound on the action, allowing us to decay the vacuum semi-classically. On the other hand, as for the string catalysis in five dimensional de Sitter spacetime, the contribution from the horizon exists. Thus, the semi-classical decay does not occur as with the case of the black hole. In this way, we found that catalytic effects by string clouds seem to depend on the dimension, so it would be interesting to explore further on this subject in various settings. 

As an application, we studied the four dimensional bubble universe proposed in \cite{Danielsson} and claimed that the black hole and the cloud of strings in the model can be seeds for the catalytic decay.  Remarkably, we found that the bubble with small positive cosmological constant, of order $\Lambda^{(4)}/M^2_4 \sim 10^{-120}$, is favored by the catalysis, by demanding appropriate scales of the black hole and the cloud of strings to reproduce the present energy densities of matter and radiation. This can be regarded as the catalytic selection of the small cosmological constant. In this analysis, we treated the tension of the bubble is a free parameter. In general, the tension depends not only on the difference between two vacua but also the shape of the potential in between. If one believes the landscape structure of vacua in string theory, a large number of tensions can be possible and the catalytic selection of cosmological constant would work well. Although, the appearance of the correct order of  the cosmological constant is quit surprising, we have to be careful that incorporation of the inflation mechanism in this scenario would change the densities of matter and radiation and may ruin the success. This issue is important but beyond the scope of this paper, so we will leave it for future work\footnote{ 
In \cite{Heckman}, the authors discussed discontinuous jumps of energy densities caused by a quantum effect in the context of F-theory and presented a model of inflation and a time-dependent equation of state for dark energy. This idea may be applicable to the issue.}.

We proposed a wrapping D-brane on a cycle with non-trivial fluxes as one of realizations of a cloud of strings in string theory. By the argument in \cite{Baryon}, non-trivial fundamental charges are induced on such D-brane, thus, the fundamental strings should end on the brane. This is nothing but the cloud of strings and can be useful because a D-brane wrapping on a cycle with fluxes is ubiquitous in string theory. It would be interesting to construct an explicit realization in the context of string theories. Also, it would be interesting to discuss the thermal effect and the information loss problem in the context of the bubble universe. The related topics were studied recently in \cite{Sasaki}. Also, It is of importance to discuss the decay of the electroweak vacuum in the standard model. See \cite{Moroi} for a precise study. Recently, in \cite{GregoryRS}, the decay of the electroweak vacuum was investigated in the brane-world scenario\footnote{Also, see \cite{PI} for a study on the higgs instability in light of the quintessence. }. It is quite interesting to study how these works fit into the present bubble universe. We would like to revisit these issues in separate publications. 

\section*{Acknowledgments}

The authors are grateful to Norihiro Tanahashi for useful discussions and Pablo Soler for explaining us the swampland conjecture and related his papers patiently. This work is supported by Grant-in-Aid for Scientific Research from the Ministry of Education, Culture, Sports, Science and Technology, Japan (No.17K05419 and No.18H01214) and Qdai-jump Research Program of Kyushu University (No.01300). The authors thank the Yukawa Institute for Theoretical Physics at Kyoto University, where the final stage of this work was done during the YITP-W-19-05 on ``Progress in Particle Physics 2019'' and YITP-W-19-10 on ``Strings and Fields 2019''.

\appendix

\setcounter{equation}{0}
\renewcommand{\theequation}{A.\arabic{equation}}
\appendix

\section{Coleman-de Luccia bounce action in $D$-dimensions}

In this appendix, we study the homogeneous nucleation of a bubble and extend the computation by Coleman and de Luccia \cite{CDL} to general dimensions. We present that the bounce action of the decay can be expressed as an analytic form by using the hypergeometric function. Consider a transition between two vacua with different cosmological constants. We denote the position of the bubble $\widetilde{R}$ and the proper time on the bubble $\tilde{\lambda}$. Both variables are normalized by \eqref{Normalize}. By putting $M_{\pm}=0$ in \eqref{tilRdot}, we get the equation for the trajectory of the bubble,
\begin{equation}
\left( {d\widetilde{R} \over d \tilde{\lambda} }\right)^2=1-\widetilde{R}^2~. \label{Rdot2CDL}
\end{equation}
One can immediately solve the equation and find that the solution is given by $\widetilde{R}=\cos \tilde{\lambda }$. Here we choose $-\pi/2\le \tilde{\lambda}\le {\pi /2}$. Substituting for \eqref{eq:8}, one can compute the on-shell bounce action
\begin{eqnarray}
  B_{CDL}=\frac{ A_{D-2}}{4\pi G_D} \left({\gamma \over \alpha} \right)^{D-2} \int^0_{-{\pi\over 2}}  d\tilde{\lambda} \widetilde{R}^{D-3} \left(\dot{\widetilde{\tau}}_+-\dot{ \widetilde{\tau}}_-\right)~,\label{CDddim}
  \end{eqnarray}
where $A_{D-2}={2\pi^{D-1 \over 2}}/\Gamma({D-1\over 2})$ is the area of $D-2$ dimensional unit sphere. As for the de Sitter space, there is the cosmological horizon that yields a singular contribution to the bounce solution, which is eventually subtracted by that of the de Sitter space. $\widetilde{\lambda}$ dependence of $\widetilde{\tau}$ is described by
\begin{equation}
\dot{\widetilde{\tau}}_{\pm}={1\over f_{\pm}}\sqrt{f_{\pm}-\left( {d\widetilde{R} \over d \tilde{\lambda} }\right)^2} ={1\over f_{\pm}}\sqrt{f_{\pm}-(1-\widetilde{R}^2)} ~.\label{tauCD}
\end{equation}
The explicit functions of $f_{\pm}$ are 
\begin{equation}
f_{\pm}(R)=1-\kappa_{\pm}\widetilde{R}^2~,
\end{equation}
where we defined 
\begin{equation}
\kappa_+=\left({\gamma \over l \alpha}\right)^2-{(\alpha^2-1)\over \alpha^2}  \ , \qquad \kappa_-={\alpha^2-1\over \alpha^2}~.
\end{equation}
Substituting for \eqref{tauCD}, we obtain
\begin{equation}
\dot{\widetilde{\tau}}_{\pm}={\widetilde{R} \over 1-\kappa_{\pm}\widetilde{R}^2 } \sqrt{1-\kappa_{\pm}}~.
\end{equation}
With these expressions and by using $\widetilde{R}$ as the integration variable, the bounce action \eqref{CDddim} reduces to 
\begin{eqnarray}
  B_{CDL}=   \frac{A_{D-2} \gamma^{D-2}}{4\pi G_D \alpha^{D-2}} \left[\sqrt{1-\kappa_+}H(\kappa_+,D)-\sqrt{1-\kappa_-}H(\kappa_-,D) \right]~,
  \end{eqnarray}
where we used the following formula,
\begin{equation}
H(\kappa, D)\equiv \int_0^1 dx {x^{D-2} \over (1-\kappa x^2)\sqrt{1-x^2}}={\sqrt{\pi} \over 2}{\Gamma \left({D-1 \over 2} \right)   \over \Gamma \left( {D \over 2} \right)} {}_2F_1\left(1,{D+1 \over 2},{D \over 2},\kappa \right)~.
\end{equation}
Here, ${}_2F_1(a,b,c,d)$ is the hypergeometric function.

%
%

\end{document}